\documentclass[conference]{IEEEtran}

\usepackage{url}
\usepackage{subfigure,times, epsfig,  graphicx, verbatim} 
\usepackage{algorithm,algorithmic}
\usepackage{psfrag}
\usepackage{dcolumn}%
\usepackage{bm}%
\usepackage{color}
\usepackage{amsmath, amssymb, latexsym,amsfonts}
\usepackage[nospace]{cite}
\usepackage{xspace}
\usepackage{multirow}
\usepackage{makecell}

\newcommand{\E}{\mathbb{E}}
\newcommand{\eqn}[1]{Eq.(#1)}

\newcommand{\Sec}[1]{Section~\ref{#1}}
\newcommand{\Fig}[1]{Fig.~\ref{#1}}
\newcommand{\Tab}[1]{Table~\ref{#1}}
\newcommand{\est}[1]{\widehat{#1}}
\newcommand{\ie}{{\em i.e., }}
\newcommand{\eg}{{\em e.g., }}

\begin{document}

\title{Estimating Clique Composition and Size Distributions from Sampled Network Data}

\author{Minas Gjoka, Emily Smith, Carter Butts\\
University of California, Irvine\\ 
{\tt \{mgjoka, emilyjs, buttsc\}@uci.edu}\\
}

\maketitle

\begin{abstract}
Cliques are defined as complete graphs or subgraphs; they are the strongest form of cohesive subgroup, and are of interest in both social science and engineering contexts.
In this paper we show how to efficiently estimate the distribution of clique sizes from a probability sample of nodes obtained from a graph (e.g., by independence or link-trace sampling). We introduce two types of unbiased estimators, one of which exploits labeling of sampled nodes neighbors and one of which does not require this information. We compare the estimators on a variety of real-world graphs and provide suggestions for their use. We generalize our estimators to cases in which cliques are distinguished not only by size but also by node attributes, allowing us to estimate clique composition by size. Finally, we apply our methodology to a sample of Facebook users to estimate the clique size distribution by gender over the social graph.

\end{abstract}

\section{Introduction}

In a large number of real-world applications it is common to represent systems, structures, or data using graphs
\eg social graphs, web graphs, or protein interaction graphs. In many cases these graphs are difficult to study, most commonly because of their massive size  and/or access limitations. As a result, there is a growing body of work \cite{Kolaczyk2009,gjoka10_walkingfb,Hardiman2009}  that uses sampling to estimate the properties of such graphs  as a step towards understanding them.  Furthermore, network models have been developed that receive as an input such estimated graph properties so as to generate synthetic graphs that resemble the real graph \cite{handcock2008statnet,gjoka13_2.5K_Graphs}. 

In this paper, we show how to efficiently estimate graph properties of the clique structure from a probability sample of nodes. Cliques are employed in a wide range of fields. In social network analysis, cliques are the foundation for studying both clustering (\ie via triangles) and cohesive subgroups \cite{wasserman1994social}. Scholars have used cliques to study advice networks in entrepreneurial firms \cite{krackhardt2002structure}, informant accuracy in behavioral and cognitive network data \cite{bernard1980informant}, and friendship networks among school classes \cite{jansson1997clique}. In bioinformatics, calculations of cliques have been used to identify protein structures \cite{grindley1993identification} and determine optimal protein structure alignment \cite{strickland2005optimal}. Other fields have studied image recognition using maximal cliques as interest points \cite{stentiford2010image} and have used maximal cliques to solve the stereo correspondence problem \cite{horaud1989stereo}. 

\begin{figure}
\centering
\includegraphics[width=0.45\textwidth]{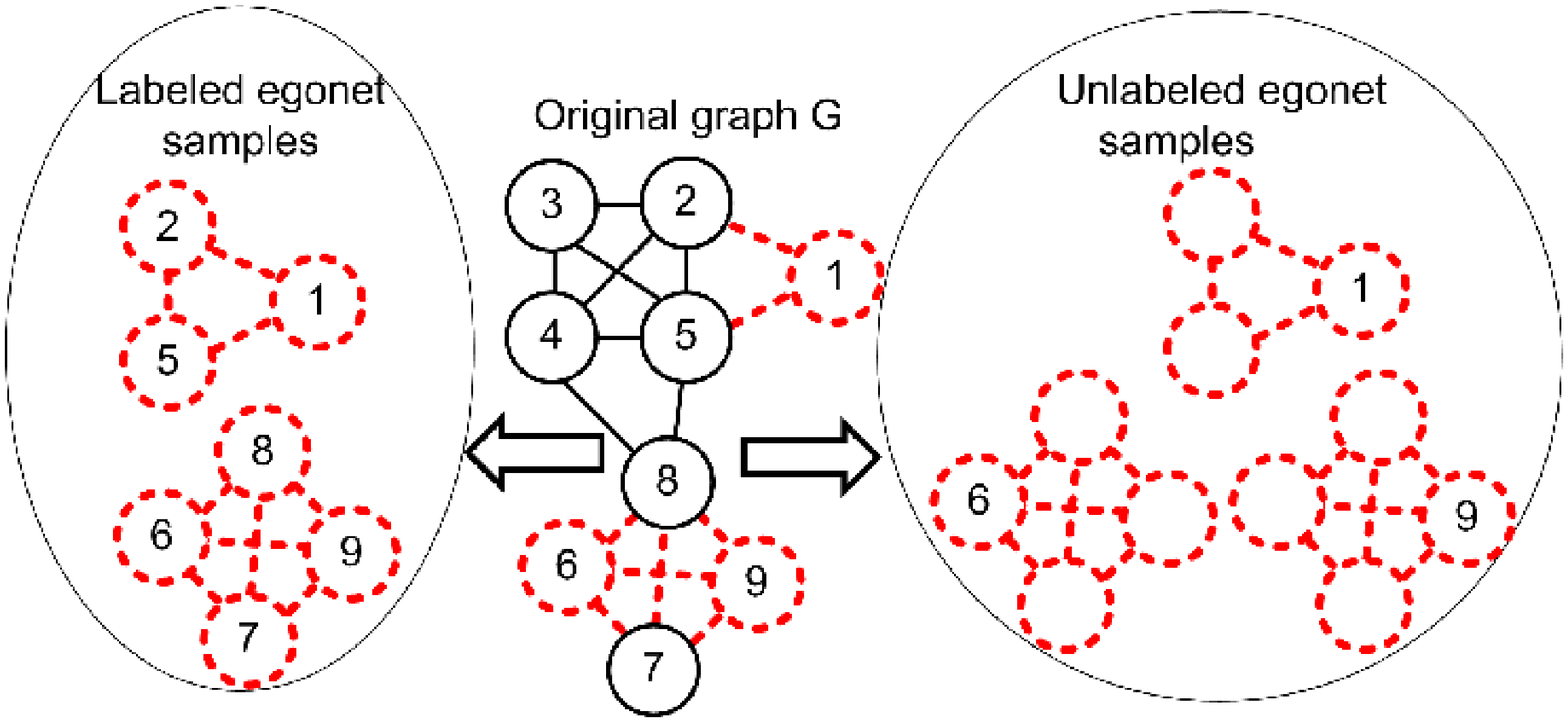}
\caption{Illustration of the egocentric approach, clique size distribution and difference between labeled and unlabeled cases. In this case, egos 1, 6, 9 are sampled from graph $G$. The maximal clique size distribution of G is $C=(0,0,2,2)$ since there are two maximal cliques of sizes $3$ and $4$.  }
\vspace{-5pt}
\label{fig:labeling}
\vspace{-10pt}
\end{figure}

Our estimation techniques employ an egocentric approach \cite{wasserman1994social}. We first collect a uniform or non-uniform probability sample of nodes (``ego'') from the target graph. Then, we collect the egonet of each sampled node, which consists of the neighbors of the node and the edges between these neighbors. \Fig{fig:labeling} presents an example that illustrates the egocentric approach. 
Next, we use an existing clique enumeration tool to calculate the exact clique distribution for each collected egonet. Last, we apply our unbiased estimators that combine the individual egonet calculations so as to estimate the clique distribution for the whole graph.

Our approach has several benefits. The first obvious benefit is that it allows us  to estimate the clique distribution of an unknown graph, as long as we have a sampling primitive that reveals the neighbors of a selected node in that graph. %
The second benefit is that it can be used to estimate the clique distributions of a fully known massive graph. Our approach decomposes a large problem into many smaller problems that can be independently computed, hence making estimation embarrassingly parallel.
Another benefit is the ability to estimate clique distribution in the absence of unique node labels \eg  due to privacy-sensitive network data or data collection limitations.  Finally, our techniques can be employed with data collected using standard techniques in both online (\eg random walk or user ID sampling \cite{gjoka2011practical}) and offline (\eg survey instruments \cite{marsden2005recent}) settings.

In summary, we make the following contributions. (i) We present two types of unbiased estimators for the clique distribution, one that exploits labeling of nodes and one that does not require this information. To the best of our knowledge, we are the first to present estimators for clique distributions in arbitrary graphs using sampled network data. We compare our estimators on a variety of real-world graphs and provide suggestions for their use. 
(ii) We generalize the clique distribution to cases in which cliques are distinguished not only by size but also by node attributes, and provide estimators for these cases. 
(iii) We apply our methodology to a sample of Facebook  (FB) users to estimate the clique size distribution and gender composition of cliques across the social graph. 

The structure of the remainder of the paper is as follows. Section~\ref{sec:related} reviews related work.
Section~\ref{sec:notation} presents the notation and objective. 
Section~\ref{sec:estimation} presents our clique distribution estimators for uniform and non-uniform probability samples.  Section~\ref{sec:simulation} presents simulation results on real-life fully known graphs. Section~\ref{sec:facebook} applies our estimators to samples collected from Facebook. 
Section~\ref{sec:conclusion} concludes the paper.

\section{Related Work}
\label{sec:related}

Egocentric sampling is a widely used method for gathering network data \cite{marsden1990network,carrington2005models}. This method samples individual 
nodes
and then expands to include their neighborhoods. 
While this procedure does not necessarily describe the structure of the entire network, it can yield representative samples of the network \cite{marsden1990network}.
Standard random sampling methods 
can be used 
to obtain egocentric network data and generalize the results to a larger population \cite{marsden1990network}. Examples of applications of egocentric sampling procedures include the network items in the General Social Survey \cite{burt1984network} and networks obtained through crawling online social networks \cite{gjoka2011practical}.

Although clique enumeration is in general exponential time in graph order (since the maximum number of cliques scales exponentially in the number of vertices), algorithms have been proposed for enumeration on small and/or sparse graphs. \cite{eppstein2011listing} introduces an algorithm to list all maximal cliques in large sparse graphs. In contrast to \cite{tomita2006worst} which uses adjacency matrices to calculate all maximal cliques, this algorithm offers an alternative for analysis of large sparse graphs whose adjacency matrices are too large for fast computation.  \cite{cheng2010finding} propose an algorithm for clique enumeration with improved space complexity, which also facilitates decomposition of clique calculations on large graphs.  All of these approaches are exact, but require processing and storage of the entire graph.

\cite{patel2010estimation} examines three methods for estimating the number of cliques in a random graph. The first is a sampling technique where they use the Bron and Kerbosch method to count a representative selection of subgraphs. This produces a curve for the graph, and they use the area under the curve to estimate the total number of cliques in the graph. The second method is using probabilistic arguments to calculate the expected number of cliques. The third is a curve fitting method. These results assume that the target graph is well-approximated by a uniform random graph. \cite{jansson1997clique} outlines two approaches for estimating non-overlapping latent clique structures in a regular graph. The first uses maximum likelihood estimates and the second uses Bayesian estimates. These techniques estimate latent (not observed) cliques, and assume both disjointness and a fixed number of edges per node.  These and other model-based approaches require assumptions regarding the network generating mechanism, 
which are avoided by our design-based approach.

\section{Notation and Objective}
\label{sec:notation}
Let $G=(V,E)$ be an 
undirected
graph with $N=|V|$ nodes and $|E|$ edges. 
For a given node $v\in V$ (``ego''), let $\mathcal{N}(v)\subseteq V$ denote the neighborhood of $v$ in $G$.  $v$'s \emph{egonet} consists of the subgraph of $G$ induced by $v$ and its neighborhood (i.e., $G[v \cup \mathcal{N}(v)]$).  In general, we will be concerned with settings in which we observe a probability sample of egonets from $G$, possibly accompanied by attributes of the associated vertices, and seek to infer properties of the clique structure of $G$.  We elaborate as follows.

\subsection{Clique Distributions}

\subsubsection{Without Attributes}
A \emph{clique} is an induced subgraph in which every vertex adjacent to every other vertex (i.e., a complete induced subgraph).  A clique that contains $i$ vertices is called an \emph{order-$i$ clique} (or \emph{$i$-clique}). A clique is said to be \emph{maximal} if no higher order clique contains it. We define $C_i$ to be the number of order-$i$ cliques in $G$, and the vector $C=(C_1,\ldots,C_N)$ the \emph{clique distribution} of $G$.  (Counts of maximal cliques, and the maximal clique distribution are defined analogously, with the constraint that the cliques in question are maximal cliques.)

\subsubsection{With Attributes}
We generalize the clique distribution to the case in which cliques are distinguished not only by size, but also by composition.  Specifically, let us assume that each vertex $j$ in the graph has some categorical attribute $X_j$; we denote the possible states of $X_j$ by the integers $1,\ldots,p$.
Potential attribute distributions within cliques are defined via \emph{composition vectors}, $u$, that indicate the number of vertices within a clique belonging to each of the $p$ categories on $X$.  Specifying $u$ tacitly specifies order (since $\sum_{j=1}^p u_j = i$, where $u_j$ is the number of vertices belonging to category $j$), and thus $u$ can be viewed as a generalization of order for purposes of clique categorization.
Accordingly, we define $C_u$ to be the number of cliques in $G$ that have composition vector $u$ (\emph{$u$-cliques}), with the set of all $C_u$ all that $u\in\{u\in\mathbb{N}^p: 1 \le \sum_{j=1}^p u_j\le N \}$ comprising the \emph{clique composition distribution} of $G$.

\subsection{Objective}
An \emph{egocentric network sample} $H_1,...,H_n$ is a probability sample of $n$ egonets from $G$. Our goal is to estimate $C_i$ and/or $C_u$ from this sample for any given order $i$ or composition vector $u$ (as defined above). When estimating $C_u$, we assume that the egocentric network sample also contains attribute information for each ego and all neighbors.

\subsection{Sample Properties}
As noted above, we assume that our egonets comprise a probability sample of the egonets in $G$; that is, (i) we can treat each ego as being included in the sample with known probability, and (ii) the probability of sampling any vertex $v$ is positive for all $v\in V$.  It is convenient to focus on two important cases:

\smallskip\noindent\textbf{Uniform Independence Sampling (UIS)}, where nodes are sampled independently with equal probabilities. 

\smallskip\noindent\textbf{Non-uniform Independence Sampling (WIS)}, where nodes are sampled independently with probability proportional to a known weight~$w(v)$.

Sampling may occur with or without replacement; we indicate these distinctions where they affect estimation.  Note also that, in practice, samples drawn using link-trace methods (e.g. \cite{gjoka2011practical}) may closely approximate UIS or WIS, and may be employed as well.  We provide an example of this approach in \Sec{sec:facebook}.

\subsection{Node Labeling}

When egonet $H_i$ of ego $i$ is sampled, it may or may not be possible to uniquely identify $i$'s neighbors (in the sense of knowing, e.g., whether $v\in H_i$ also belongs to some $H_j$).  When such identification is possible, we say that the sample is \emph{labeled}, otherwise denoting it as \emph{unlabeled}. \Fig{fig:labeling} shows the effect of labeling in an example graph in which egos 1,6, and 9 (center graph, red) are sampled.  If the sample is labeled (left), we can discern that the three sampled nodes belong to two maximal cliques: $\{1,2,5\}$ and $\{6,7,8,9\}$.  In the unlabeled case (right), however, we know only that ego 1 belongs to a maximal 3-clique, and that egos 6 and 9 each belong to one maximal 4-clique.  As we will show, estimation is possible in both cases; however, labeled samples provide additional information that can be leveraged to reduce sampling error.

\section{Estimation}
\label{sec:estimation}

As stated, our goal is to infer $C_i$ and/or $C_u$ for some set of $i$ and/or $u$ from an egocentric sample.  Our methods are applicable to either cliques or maximal cliques, so long as the same definition is applied consistently throughout.  In what follows, we will assume that a sample of $n'$ egos was taken from the population of vertices, of which $n$ are unique; in the case of sampling with replacement, $n$ may be less than $n'$.  Our data consists of $n'$ and $n$ together with the sample $H_1,\ldots,H_n$ of egonets associated with unique egos (i.e., any repeatedly sampled egonets appear only once).  Note that it is still possible that vertices may appear in multiple egonets, a useful fact that we exploit.

\subsection{General Techniques}

We use two basic techniques to estimate $C_i$ and $C_u$, one of which can be used with either labeled or unlabeled neighborhoods and the other of which requires labeling information.  We present these general techniques here, followed by specific details on how they may be used for estimation in different sampling and/or labeling scenarios.

\paragraph{Estimation via Clique Degree Sums (CDS)}

In this approach, we transform the problem of estimating $C_i$ or $C_u$ into one of estimating a sum of local structural features, and apply standard (Horvitz-Thompson) sampling theory to obtain an estimate.  The approach does not require labeled neighborhoods.  We begin by defining the \emph{order-i clique degree} and \emph{u-clique degrees} of node $j$, denoted $d_{ij}$ and $d_{uj}$, to be respectively the numbers of $i$-cliques and $u$-cliques to which $j$ belongs.  Totaling these quantities over $V$ gives us the corresponding \emph{clique degree sums},
\begin{equation}
D_i = \sum_{j=1}^N d_{ij}  \quad \textrm{and} \quad D_u = \sum_{j=1}^N d_{uj},
\end{equation}
of which the conventional degree sum $D_2$ is a special case.  The clique degree sums are important because of their linear relationships with $C_i$ and $C_u$ respectively.  In particular, note that every $i$-clique appears $i$ times in $D_i$; since this is true irrespective of composition, we take without loss of generality $i=\sum_j u_j$ for the $u$-clique case to simplify exposition.  From this it immediately follows that $D_i=i C_i$ and $D_u=i C_u$, and hence
\begin{equation}
C_i = D_i/i  \quad \textrm{and} \quad C_u = D_u/i.
\end{equation}

Since $i$ is a known constant, this gives us a direct means of estimating the clique counts: for any unbiased estimators $\est{D_i}$ and $\est{D_u}$ of $D_i$ and $D_u$, the estimators $\est{C_i}=\est{D_i}/i$ and $\est{C_u}=\est{D_u}/i$ are unbiased estimators of $C_i$ and $C_u$ respectively. \footnote{An unbiased estimator, $\est{t}$ of $t$ is a statistic such that $\E[\est{t}]=t$. The result follows immediately from the linearity of expectation.}  Moreover, it also follows from standard properties of the variance that $Var(\est{C_i})=Var(\est{D_i})/i^2$ and $Var(\est{C_u})=Var(\est{D_u})/i^2$ since we are merely multiplying a random variable by a constant.  Thus, the sampling variability of our clique count estimators can be determined immediately from the properties of our degree sum estimators.

To estimate clique degree sums from egonet data, we employ Horvitz-Thompson (H-T) estimators \cite{horvitz.thompson:jasa:1952}, which are design-unbiased (with or without replacement) and which are asymptotically Gaussian for many data collection designs \cite{thompson:bk:2002}.  Define $p_j$ as the a priori probability of $H_j$ appearing in the sample (i.e., the $j$th unique ego having been sampled at any point), and let $d_{i}(H_j)$ and $d_u(H_j)$ be respectively the order-$i$ clique degree and $u$-clique degree of ego $j$.  The H-T estimators of $D_i$ and $D_u$ are then
\begin{equation}
\est{D_i} = \sum_{j=1}^n \frac{d_{i}(H_j)}{p_j}  \quad \textrm{and} \quad \est{D_u} = \sum_{j=1}^n \frac{d_{u}(H_j)}{p_j},
\end{equation}
with the H-T estimates of $C_i$ and $C_u$ obtained by dividing the degree sum estimators by $i$.  Note that since $j$'s ego net contains all of its cliques, $d_i$ and $d_u$ require only local network information; this can also greatly speed computation, as discussed in \Sec{subsec:implementation}.  Since we are concerned only with ego's clique membership, we do not require labeled neighborhoods for this method nor do our results require any assumptions regarding possible neighborhood overlap.  The inclusion probabilities for sample elements arise from the design, and may differ for distinct $j$.  We discuss this further below.

For designs where the probability that any two observed nodes, $j$ and $k$, are both included in the sample is known, unbiased estimators of the clique count estimator variance are given by the general form \footnote{This is a direct application of Eq. 6 of \cite{thompson:bk:2002}, p54.}
\begin{equation}
\begin{split}
\est{Var}(\est{C_*}) =& \sum_{j=1}^n \left(\frac{1}{p_j^2}-\frac{1}{p_j}\right)\left(\frac{d_{*}(H_j)}{i}\right)^2\\
& + 2 \sum_{j=1}^n \sum_{k=j+1}^n \left(\frac{1}{p_jp_k}-\frac{1}{p_{jk}}\right)\left(\frac{d_{*}(H_j)d_{*}(H_k)}{i^2}\right),
\end{split}
\end{equation}
where $*$ is replaced with the desired order ($i$) or composition vector ($u$) as appropriate.  $p_{jk}$ above is the probability of both $j$ and $k$ appearing in the sample.  
For designs such that $p_{jk}$ cannot be readily determined, the generalized H-T estimators of form $\eqn{\ref{e_genestvar}}$ below can be employed.

An important special case arises when sample inclusion probabilities are unequal and known only up to a constant factor (i.e., some $w_j \propto p_j$).  In this case, generalized H-T estimators \cite{thompson:bk:2002} are required for estimation of clique counts:
\begin{equation}
\est{C_i} = \frac{N}{i}\frac{d_{i}(H_j)/w_j}{\sum_{j=1}^n 1/w_j} \quad \mathrm{and} \quad \est{C_u} = \frac{N}{i}\frac{d_{u}(H_j)/w_j}{\sum_{j=1}^n 1/w_j}.
\end{equation}
The denominator in these estimators may be recognized as an H-T estimate of the total population weight (and hence a normalizer of the weights $w_j$).  The above estimators are asymptotically unbiased \cite{thompson:bk:2002}; given that joint inclusion probabilities are not available here, an adaptation of the Brewer and Hanif (B-H) variance estimator \cite{brewer.hanif:bk:1983} leads to the following general form:
\begin{equation}
\est{Var}(\est{C_*}) = \left(\frac{N-n}{n(n-1)N }\right)\sum_{j=1}^n\left(\frac{n d_*(H_j)/w_j}{\sum_{k=1}^n i/w_k}-\est{C_*}\right)^2 
\label{e_genestvar} 
\end{equation}
(where $*$ is replaced with $i$ or $u$ as appropriate).  The B-H estimator is generally biased upward \cite{thompson:bk:2002}, and is hence a conservative estimate of measurement error, but does not require joint inclusion weights.

\paragraph{Estimation via Distinct Clique Counting (CC)}

In the case of labeled neighborhoods, we can identify particular cliques across egonets.  This allows us to develop H-T estimates for clique counts that do not require degree sums.  Let $c'_i$ be the number of \emph{distinct} $i$-cliques in $H_1,\ldots,H_n$ that contain the egos of their respective egonets, with $c'_u$ the corresponding number of distinct $u$-cliques.  We define $\pi_k$ to be the probability of the $k$th such clique appearing in the sample (i.e., the probability of selecting at least one of its members as an ego).  Given $\pi_k$, H-T estimators of the clique counts are given by
\begin{equation}
\est{C_i} = \sum_{k=1}^{c'_i} \frac{1}{\pi_k} \quad \mathrm{and} \quad \est{C_u} = \sum_{k=1}^{c'_u} \frac{1}{\pi_k}
\end{equation}
(with the $\pi$ values in each sum being specific to that set of cliques).  Because joint sampling probabilities of cliques depend on the structure of clique overlap (and are not in general known), we propose to estimate the variance of the above via the B-H estimators
\begin{equation}
\est{Var}(\est{C_*}) = \left(\frac{\est{C_*}-c'_*}{c'_*(c'_*-1)\est{C_*} }\right)\sum_{k=1}^{c'_*}\left(\frac{n c'_* }{\pi_k}-\est{C_*}\right)^2,
\end{equation}
with $*$ replaced by $i$ or $u$ as appropriate.  Intuitively, the above estimates the variability in count estimates associated with each sampled clique, and rescale the result to reflect the estimate of the effective sample size (i.e. number of cliques observed) relative to the total population.

\subsection{Inclusion Probabilities}

We have provided estimators (and variance estimators) for clique counts based on order and/or composition for either labeled or unlabeled egonet samples.  To use them, it remains only to determine the inclusion probabilities of nodes or cliques ($p$ and $\pi$, above).  These quantities depend on the sampling design; we here provide examples for some common and important cases for sampling of OSNs in particular but also other arbitrary graphs.

\paragraph{Node inclusion probabilities}

The simplest case for node inclusion probabilities is that in which egos are sampled uniformly at random from the population (UIS).  The inclusion probabilities depend upon the total number of samples drawn ($n'\ge n$), and whether samples are drawn with or without replacement.  In the with-replacement case, an arbitrary node $j$ is fails to be selected on any given draw with probability $1-1/N$, and hence is ultimately included with total probability $p_j=1-(1-1/N)^{n'}$.  When sampling is performed without replacement, a total of $n'=n$ of the $N$ available nodes are drawn, any of which could be $j$.  The resulting inclusion probability is thus simply $p_j=n'/N$.

Under UIS, joint inclusion probabilities for nodes are also easily determined.  For arbitrary nodes $j,k$ under with-replacement UIS, the probability of observing both $j$ and $k$ in the sample is given by $p_{jk}=1-2((N-1)/N)^{n'}+((N-2)/N)^{n'}$; without replacement, the corresponding probability is $p_{jk}=n(n-1)/(N(N-1))$.  Both arise from standard combinatorial arguments.

When the probability of inclusion on any given draw is unequal, total inclusion probabilities may depend on the details of the sampling mechanism.  In the common case of independent with-replacement sampling with unequal probabilities (WIS), the probability of including node $j$ can be determined from the probability of obtaining $j$ on any \emph{given} draw, $p'_j$, by $p_j=1-(1-p'_j)^{n'}$.  Without-replacement inclusion probabilities with unequal are not easily summarized, but computational tools such as \cite{tille.matei:sw:2012} can be employed to obtain them.

In some cases the per-draw inclusion probability may be unequal and known only up to a constant factor (i.e., $w'_j \propto p'_j$).  This situation is common in e.g. random walk sampling of OSNs, where vertices are often sampled (approximately) independently with replacement, proportional to degree.  In such cases, approximating $p'_j$ by the Hansen-Hurwitz estimator $p'_j \approx w'_j \left(\sum_{k=1}^{n'} 1/w'_k\right)/(n'N)$ (where the sum is over all observations, including repetitions) is a practical alternative.

\paragraph{Clique inclusion probabilities}

The probability of sampling a clique is equal to the probability of sampling at least one of its members.  In the with-replacement UIS case, this is simply $\pi_k=1-(1-i/N)^{n'}$, where $i$ is the order of the clique.  When nodes are drawn uniformly without replacement, the corresponding inclusion probability becomes
\[
\pi_k =1- \prod_{k=0}^{n'} \frac{N-i-k}{N-k}
\]
(i.e., one minus the chance of sequentially drawing $n$ nodes from the population of non-clique members).

In the WIS case, clique inclusion probabilities will vary with the inclusion probabilities of their members.  For the $k$th clique in an $i$ or $u$-defined set, let $m_k$ be the $i$-vector denoting the clique members.  The resulting inclusion probability is then $\pi_k = 1-\left(1-\sum_{j=1}^i p'_{m_{kj}}\right)^{n'}$, where $p'$ contains the per-draw node sampling probabilities.  More complex unequal probability designs (e.g., without replacement) do not allow simple specification of clique inclusion probabilities, but as with the nodal case these may be computed in particular cases.  Since the approach to be used depends on details of the sampling procedure, we do not consider this further here.

\subsection{Implementation considerations}
\label{subsec:implementation}
The estimation of clique distributions requires the graph size $N$ and the enumeration of cliques for each sampled egonet. We point out that in the cases when $N$ is not known a priori, \cite{Katzir2011} provides estimators that work with sampled network data. In general, the enumeration of maximal cliques is an NP-hard problem with worst case complexity $O(3^{N/3})$. In our approach we decompose the enumeration over the whole graph to enumeration for each sampled egonet separately. Since an egonet has in the worst case size $D$ equal to the maximum degree, the worst case complexity is $O(n*3^{(D/3)})$ where $n$ is the egonet sample size; this can be trivially reduced to $O(3^{(D/3)})$ via parallel computation on each egonet. It is worth noting that in real-world graphs there is often a large difference between the graph size $N$ and maximum degree $D$ \eg in the FB social graph $N=1.11$ billion whereas $D=5000$. Further, our method is flexible to support partial enumeration of cliques up to order-$i$.

The space complexity of the Clique Degree Sum method, which works only with unlabeled needs, is $O(len(C)))$ which is almost negligible. On the other hand, the distinct Clique Counting methods requires $O(\sum_{i=1} C_i)$) which can be quite large depending on the graph. In \Sec{sec:simulation} we describe a heuristic that helps us choose between CDS and CC.

\section{Performance Evaluation via Simulated Sampling}
\label{sec:simulation}

In this section, we evaluate the performance of the estimators with labeled and unlabeled neighborhoods (i.e., direct counting versus clique degree sums) via simulated uniform without-replacement sampling from real-world datasets.  Our results shed light on the relative advantages of these estimators for assessing both order and composition distributions.

\begin{table}[h]
\centering
{
\begin{tabular}{|@{}r@{}|@{}r@{}|@{}r@{}|@{}r@{}|@{}r@{}|@{}r@{}|@{}r@{}|}
\hline
    Dataset          & $|V|$   & $|E|$ & \multicolumn{1}{@{}c@{}|}{Average} &  \multicolumn{1}{@{}c@{}|}{Max}   &  \multicolumn{1}{@{}c@{}|}{\# Maximal}  & \multicolumn{1}{@{}c@{}|}{Maximum} \\
                     &         &       & \multicolumn{1}{@{}c@{}|}{Degree} &  \multicolumn{1}{@{}c@{}|}{Degree} &  \multicolumn{1}{@{}c@{}|}{Cliques} & \multicolumn{1}{@{}c@{}|}{Clique Size} \\
\hline
FB:New Orl.~\cite{Viswanath2009}                 &       63\,392  &     816\,884 &  25.77 & 1\,098 &   1\,538\,105  & 30\\
email-EuAll~\cite{WWW_SNAP_Graph_Library}        &     224\,832   &   339\,923\, &  3.02  & 7\,636 &    353\,194  & 16\\
soc-sign-Epin~\cite{WWW_SNAP_Graph_Library}      & \xspace \xspace \xspace 119\,129       &  \xspace \xspace \xspace 704\,265     & 11.82  & 3558  & 22\,219\,084  &  94\\
soc-Slashdot~\cite{WWW_SNAP_Graph_Library}       &     77\,360    &     469\,179 &  12.13 & 2\,539 &    823\,412  & 26\\
amazon0601~\cite{WWW_SNAP_Graph_Library}         &     403\,364   &  2\,443\,309 &  12.11 & 2\,752 & 1\,023\,558  & 11\\
roadnet~\cite{WWW_SNAP_Graph_Library}            &  1\,087\,561   &  1\,541\,512 & 2.83   & 9      & 1\,413\,059  & 4\\
ca-CondMat~\cite{WWW_SNAP_Graph_Library}         &     21\,362    &     91\,282  &  8.55  & 279    &     17\,757  & 26\\
web-Google~\cite{WWW_SNAP_Graph_Library}         &  855\,802      & 4\,291\,350  & 10.02  & 6\,332 & 1\,401\,600  & 44 \\
youtube-links\cite{Mislove2007}                  & 1\,134\,894    &  2\,987\,623 & 5,26   & 28\,754& 3\,265\,955  & 17\\
\hline
FB:Duke~\cite{Traud2011}                        &    9\,895      & 506\,442     & 102.36 & 1\,887  & 8\,474\,776  &34\\ 
FB:UVA~\cite{Traud2011}                         &       17\,196  & 789\,321     & 91.80  & 3\,182  & 4\,727\,791  & 42\\
FB:UCSD~\cite{Traud2011}                        &      14\,948   &     443\,221 &  59.30 & 2\,165  &  743\,328  & 43\\
\hline
\end{tabular}}
\caption{Empirical topologies used in Sec.~\ref{sec:simulation} %
}
\vspace{-20pt}
\label{tab:Topologies}
\end{table}

\subsection{Datasets}
\Tab{tab:Topologies} lists the empirical networks that we use in our evaluation study. It includes several online social networks, an email communication graph, a co-authorship network, a transportation topology and a web graph; we here treat all structures as undirected graphs. For each network we list the number of nodes, number of edges, average and maximum degree,  sum of maximal cliques over the whole distribution, and the maximum clique size.  The numbers of nodes and sums of maximal cliques in the graphs range from thousands to millions.

The list is composed of two groups. The first group contains no attributes and will be used to estimate $C_i$. The second group contains several node attributes. We have selected the node attribute ``gender'' to estimate $C_u$.

\subsection{Error Metric}
We measure the difference between estimated and actual clique distributions using the Normalized Mean Absolute  Error (NMAE), defined as:
\begin{equation}
\label{eq:NMAE}
NMAE(\est{\vec{x}},\vec{x}) = \frac{ \sum(|\est{x}_i-x_i|)  }{\sum |x_i|},
\end{equation}
where~$\vec{x}$ and $\est{\vec{x}}$ are the vectors that correspond to the real and estimated distributions. $NMAE$ returns the absolute estimation error relative to the true value, averaged over every point in the distribution. 

\subsection{Results}

\begin{figure}
\centering
\subfigure[NMAE distribution for sample sizes ($n$) 125--32,000]{
\includegraphics[width=0.48\textwidth]{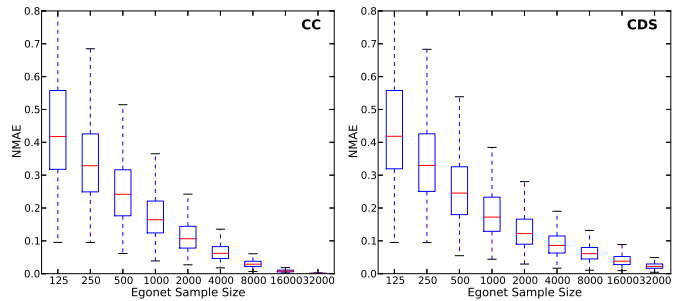}
}
\subfigure[Distribution of simulated $\est{C_i}$ versus true values; $n=1000$, $i\in 1,\ldots,30$]{
\includegraphics[width=0.48\textwidth]{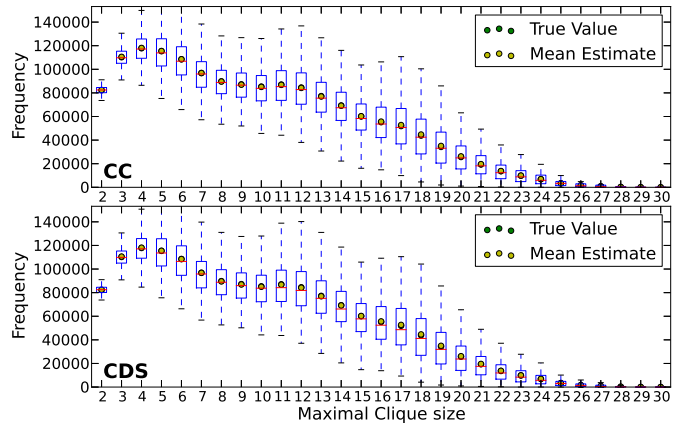}
\label{fig:neworleans_Ci_order}
}
\caption{(FB: New Orleans) Clique size distribution ($C_i$) estimates for 1000 simulated data sets.  Egonets sampled uniformly without replacement.}
\vspace{-10pt}
\label{fig:neworleans_Ci}
\end{figure}

\begin{figure*}
\centering
\includegraphics[width=0.9\textwidth]{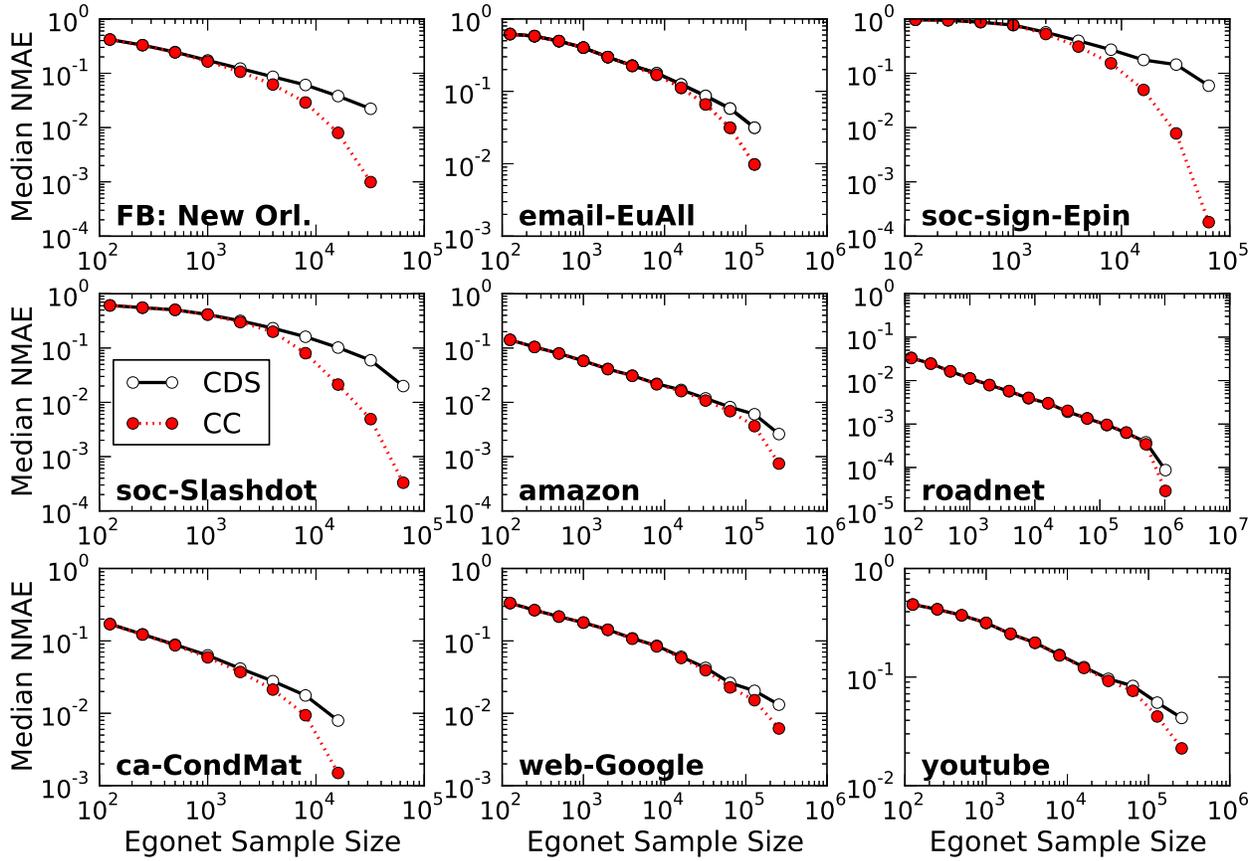}
\vspace{-10pt}
\caption{Clique order distribution for real-world topologies. Median NMAE for the estimation of $C_i$ calculated over 1000 replications, as a function of $n$ (uniform sampling without replacement).  }
\vspace{-10pt}
\label{fig:allCi_nmae}
\end{figure*}

\begin{figure*}
\centering
\includegraphics[width=0.9\textwidth]{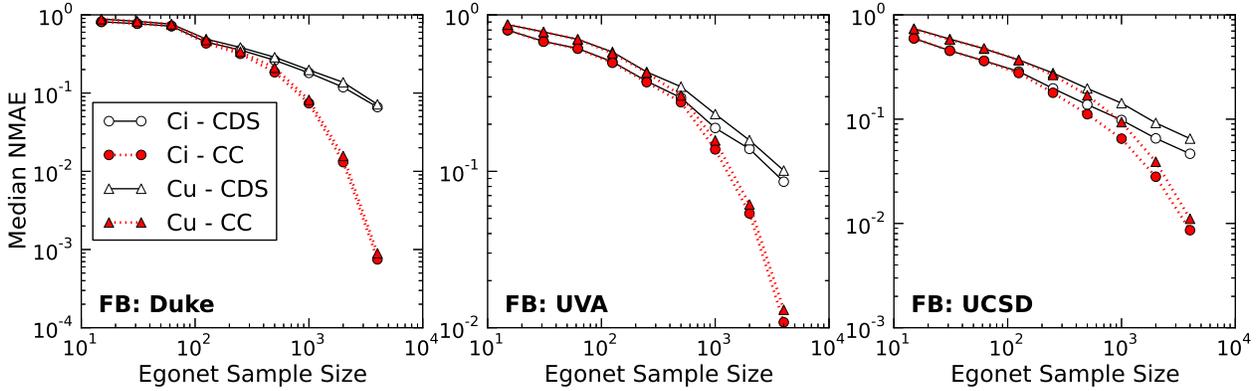}
\vspace{-10pt}
\caption{Clique order and composition distributions for real-world topologies. Median NMAE for the estimation of $C_i$ and $C_u$ over 1000 replications, as a function of $n$ (uniform sampling without replacement). }
\vspace{-10pt}
\label{fig:allCi_Cu_nmae}
\end{figure*}

\Fig{fig:neworleans_Ci} shows the simulation results for the topology ''FB: New Orleans'' and egonet sample sizes ($n$) 125-32\,000. We observe that with only 125 egonets the median NMAE error for both  distinct Clique Counting (CC) and Clique Degree Sum (CDS) estimators is already fairly small at $\sim 0.42$. As the sample size increases, the error rapidly declines (dropping below 10\%, on average, for $n\approx 4e3$). In \Fig{fig:neworleans_Ci_order} we ``zoom in'' to the case of $n=1000$ to observe the sampling distribution of estimates for all order-$i$ cliques in the distribution.  Confirming the unbiasedness of our estimators, the mean estimates (and in this case the median estimates) closely match the true values for all observed clique orders (1--30).

In \Fig{fig:allCi_nmae} we plot the median NMAE of the CC and CDS estimators of $C_i$ for various real-world topologies as a function of sample size. 
We vary $n$ from $125$ to the total size of each graph, allowing us to observe the effects of saturation on measurement error.  We note that for smaller sample sizes, the CC and CDS estimators perform equally well.  Beyond a threshold sample size, however, the CC begins to substantially outperform the CDC estimator (reflecting the additional information associated with vertex labels). We use \Tab{tab:results} to better interpret these results and shed some light on the causes of the ``threshold'' behavior. \Tab{tab:results} contains for each topology and egonet sample size the average \% of all nodes and \% of all edges sampled when both egos and neighbors are included. We observe that the CC ``breakaway'' threshold varies for different graphs even when taking into account the total \% of nodes and edges sampled. As an example, the threshold for the network ''soc-sign-Epinions`` is at $n\approx$4\,000, corresponding to $\approx$ 18.1\% of all nodes being sampled and 34.8\% of all edges being contained in 
some egocentric network sample on average (over 1000 simulations). On the other hand, the threshold for the network ``amazon0601'' is at $n=$64\,000, at which point 80.9\% of all nodes and 63.3\% of all edges have been captured by some egocentric sample on overage.  While saturation aids the CC estimator relative to the CDS estimator, the degree of saturation required varies markedly.

\Fig{fig:allCi_Cu_nmae} shows the median NMAE error of the CC and CDS estimators of $C_i$ and $C_u$ for several empirical networks with vertex attributes. Due to the size and density of these topologies, the egonet sample size is set between $15-4\,000$. \Tab{tab:results} shows the values for the mean \% of nodes and \% of edges sampled for these egonet sample sizes. As expected from the larger number of values (and smaller counts), estimation of $C_u$ is at least as hard as the estimation of $C_i$. Depending on the composition of the attributes, the estimation of $C_u$ ranges from being indistinguishable from $C_i$ (see FB:Duke) or slightly worse than $C_i$ (see FB:UCSD). 

Our results show clear returns to the use of labeled neighborhoods where possible: the CC estimators perform as well or better than the CDS estimators in all cases.  However, to count the distinct cliques the CC estimator needs additional space as discussed in \Sec{subsec:implementation}.
Depending on the topology, the amount of space required to implement the CC estimator might be considerably high. For example, the estimation of the clique distributions with labeling for the Facebook '09 data samples in \Sec{sec:facebook} requires space that is in the order of hundreds of GB. One question that naturally rises is whether there is a heuristic that can suggest whether the CC estimator is worth applying by a preliminary analysis of egonet data (\ie before we do any clique enumeration and estimation whatsoever).  Two such heuristics that we examined are the use of average edge count and average node count as indicators of saturation. They are defined and included in \Tab{tab:results} for all networks and egonet sample sizes 
examined here. 

\begin{figure}
\centering
\includegraphics[width=0.48\textwidth]{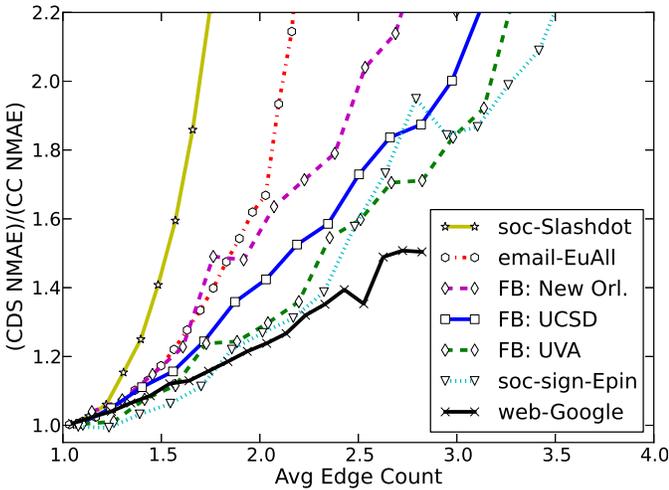}
\vspace{-5pt}
\caption{Ratio of the error between CDS and CC estimators as a function of Average Edge Count i.e. a ratio of 1.30 means that the NMAE error for the CC estimator was 30\% smaller than the CDC estimator.  Larger values indicate greater returns to use of node labels.
}
\vspace{-10pt}
\label{fig:avgedgecount}
\end{figure}

\Fig{fig:avgedgecount} shows the ratio of the error between CDS and CC estimators as a function of the  average edge count. (Values greater than 1 favor CC.)  Empirically, we observe  that a value of Average Edge Count above $1.5$ is a heuristic indicator that a CC clique estimator for a specific egonet sample significantly outperforms the corresponding CDS clique estimator. Intuitively, the average edge count of an egonet sample is correlated with the percentage of cliques that are not distinct. The higher the number of non-distinct cliques the more information the CC estimator uses compared to the CDS estimator.

Of course, in some settings (e.g., due to privacy or data collection limitations, particularly offline) is not possible to obtain information on neighbors' identities.  In these cases, our simulations suggest that the CDS estimator can still provide excellent performance, even for very large graphs.

\begin{table}
\scriptsize
  \centering
  {
\begin{tabular}{|@{}l@{}|@{}r@{}|@{}r@{}|@{}r@{}|@{}r@{}|@{}r@{}|@{}r@{}|@{}r@{}|@{}r@{}|@{}r@{}|@{}r@{}|}
\hline
     Dataset \space                    & Property            &            \multicolumn{9}{c|}{Egonet Sample Size}  \\  \cline{3-11} 

                               &                   &  \xspace \xspace  \xspace  \textbf{250}   & \xspace \xspace \xspace \textbf{500}    &  \xspace \xspace \xspace  \xspace \textbf{1K}   &  \xspace \xspace \xspace \xspace \textbf{2K}      & \xspace \xspace  \xspace \xspace \textbf{4K}     & \xspace \xspace \xspace \xspace \textbf{8K}     &  \xspace \xspace \xspace  \textbf{16K}   & \xspace \xspace \xspace  \textbf{32K} &  \xspace \xspace \xspace  \textbf{64K}  \\ 
\hline
\multirow{4}{*}{\begin{minipage}{0.1in}FB: New Orl.\end{minipage}}     
&   \% nodes sampled & 9.11  &  16.16  &  26.61  &  40.04  &  55.25  &  70.56  &  84.53  &  95.38  &  -  \\
&  \% edges sampled & 5.53  &  10.55  &  19.08  &  32.12  &  49.33  &  67.82  &  84.27  &  95.75  &  -  \\
&   avg edge count & 1.06  &  1.11  &  1.23  &  1.45  &  1.90  &  2.76  &  4.45  &  7.83  &  -  \\
&  avg node count & 1.16  &  1.31  &  1.59  &  2.11  &  3.06  &  4.78  &  8.00  &  14.17  &  -  \\

\hline \hline

\multirow{4}{*}{\begin{minipage}{0.1in}email-EuAll\end{minipage}}     
&   \% nodes sampled & 0.38  &  0.76  &  1.37  &  2.53  &  4.65  &  8.75  &  16.35  &  30.10  &  52.66  \\
&  \% edges sampled & 0.47  &  0.97  &  1.86  &  3.58  &  6.70  &  12.25  &  21.61  &  36.49  &  58.66  \\
&   avg edge count & 1.01  &  1.02  &  1.04  &  1.07  &  1.14  &  1.25  &  1.43  &  1.70  &  2.13  \\
&  avg node count & 1.16  &  1.24  &  1.34  &  1.44  &  1.55  &  1.65  &  1.76  &  1.91  &  2.18  \\

\hline \hline

\multirow{4}{*}{\begin{minipage}{0.1in}soc-sign-epinions\end{minipage}}     
&   \% nodes sampled & 2.23  &  4.05  &  6.86  &  11.31  &  18.14  &  28.61  &  43.77  &  64.57  &  87.92  \\
&  \% edges sampled & 4.31  &  8.36  &  14.21  &  23.30  &  34.84  &  49.10  &  64.32  &  80.14  &  93.97  \\
&   avg edge count & 1.07  &  1.16  &  1.33  &  1.65  &  2.19  &  3.15  &  4.78  &  7.72  &  13.10  \\
&  avg node count & 1.19  &  1.34  &  1.56  &  1.90  &  2.36  &  3.02  &  3.93  &  5.34  &  7.83  \\

\hline \hline

\multirow{4}{*}{\begin{minipage}{0.1in}soc-slashdot\end{minipage}}        
&   \% nodes sampled & 3.60  &  6.51  &  11.15  &  18.31  &  28.71  &  43.14  &  61.75  &  82.79  &  98.83  \\
&  \% edges sampled & 1.68  &  3.31  &  6.20  &  11.38  &  19.90  &  33.39  &  53.15  &  77.95  &  98.43  \\
&   avg edge count & 1.03  &  1.06  &  1.13  &  1.24  &  1.43  &  1.71  &  2.15  &  2.93  &  4.65  \\
&  avg node count & 1.17  &  1.30  &  1.52  &  1.85  &  2.36  &  3.14  &  4.40  &  6.56  &  10.99  \\

\hline \hline

\multirow{4}{*}{\begin{minipage}{0.1in}amazon\end{minipage}}      
&   \% nodes sampled & 0.80  &  1.60  &  3.15  &  6.15  &  11.72  &  21.59  &  37.48  &  59.07  &  80.95  \\
&  \% edges sampled & 0.42  &  0.85  &  1.69  &  3.35  &  6.56  &  12.60  &  23.37  &  40.59  &  63.35  \\
&   avg edge count & 1.00  &  1.01  &  1.01  &  1.02  &  1.04  &  1.08  &  1.17  &  1.35  &  1.73  \\
&  avg node count & 1.01  &  1.02  &  1.03  &  1.06  &  1.11  &  1.20  &  1.39  &  1.76  &  2.57  \\

\hline \hline 
\multirow{4}{*}{\begin{minipage}{0.1in}roadnet PA\end{minipage}}     
&   \% nodes sampled & 0.09  &  0.18  &  0.35  &  0.70  &  1.40  &  2.79  &  5.51  &  10.78  &  20.60  \\
&  \% edges sampled & 0.05  &  0.10  &  0.20  &  0.39  &  0.78  &  1.56  &  3.11  &  6.16  &  12.10  \\
&   avg edge count & 1.00  &  1.00  &  1.00  &  1.00  &  1.00  &  1.00  &  1.01  &  1.02  &  1.04  \\
&  avg node count & 1.00  &  1.00  &  1.00  &  1.00  &  1.01  &  1.01  &  1.02  &  1.05  &  1.10  \\

\hline \hline
\multirow{4}{*}{\begin{minipage}{0.3in}ca-condmat\end{minipage}\xspace}     
&   \% nodes sampled & 10.02  &  18.35  &  31.61  &  50.08  &  71.39  &  89.91  &  99.25  &  -  &  -  \\
&  \% edges sampled & 8.43  &  16.04  &  29.04  &  48.47  &  71.87  &  91.32  &  99.53  &  -  &  -  \\
&   avg edge count & 1.06  &  1.11  &  1.23  &  1.47  &  1.99  &  3.13  &  5.74  & -  &  -  \\
&  avg node Count & 1.11  &  1.22  &  1.42  &  1.79  &  2.50  &  3.98  &  7.21  &  -  & -  \\

\hline \hline
\multirow{4}{*}{\begin{minipage}{0.1in}web-google\end{minipage}}     
&   \% nodes sampled & 0.32  &  0.62  &  1.23  &  2.39  &  4.61  &  8.69  &  15.85  &  27.40  &  43.95  \\
&  \% edges sampled & 0.33  &  0.65  &  1.31  &  2.57  &  5.02  &  9.61  &  17.79  &  31.05  &  49.71  \\
&   avg edge count & 1.00  &  1.01  &  1.02  &  1.03  &  1.06  &  1.10  &  1.19  &  1.37  &  1.71  \\
&  avg node count & 1.02  &  1.03  &  1.05  &  1.08  &  1.12  &  1.19  &  1.30  &  1.51  &  1.88  \\

\hline \hline

\multirow{4}{*}{\begin{minipage}{0.2in}youtube links\end{minipage}}         
&   \% nodes sampled & 0.13  &  0.26  &  0.52  &  0.98  &  1.88  &  3.51  &  6.46  &  11.65  &  20.28  \\
&  \% edges sampled & 0.10  &  0.21  &  0.46  &  0.85  &  1.71  &  3.27  &  6.19  &  11.48  &  20.15  \\
&   avg edge count & 1.00  &  1.01  &  1.01  &  1.02  &  1.05  &  1.08  &  1.14  &  1.24  &  1.40  \\
&  avg node count & 1.03  &  1.05  &  1.08  &  1.12  &  1.18  &  1.26  &  1.36  &  1.52  &  1.73  \\

\hline 
\hline

    Dataset & Property             &            \multicolumn{9}{c|}{Egonet Sample Size}  \\
  \cline{3-11} 
            &                       &  \xspace  \xspace  \textbf{15}   & \xspace \xspace \textbf{31}    &  \xspace  \xspace \textbf{62}   &  \xspace \xspace  \textbf{125}      & \xspace  \xspace  \textbf{250}     & \xspace \xspace \textbf{500} & \xspace \xspace  \textbf{1K}   & \xspace  \textbf{2K} & \xspace  \textbf{4K}  \\

\hline 
\multirow{4}{*}{\begin{minipage}{0.1in}FB: Duke\end{minipage}}     
&   \% nodes sampled & 13.39  &  25.06  &  41.12  &  58.86  &  74.26  &  84.65  &  91.20  &  95.44  &  98.35    \\
&  \% edges sampled & 4.57  &  9.27  &  17.65  &  30.88  &  49.43  &  69.09  &  85.05  &  94.71  &  98.84    \\
&   avg edge count & 1.03  &  1.08  &  1.16  &  1.33  &  1.67  &  2.39  &  3.83  &  6.91  &  13.30    \\
&  avg node count & 1.12  &  1.28  &  1.58  &  2.22  &  3.53  &  6.20  &  11.41  &  21.86  &  42.52   \\
 
\hline \hline
\multirow{4}{*}{\begin{minipage}{0.1in}FB: UVA\end{minipage}}     
&   \% nodes sampled & 7.42  &  14.50  &  25.66  &  41.60  &  60.11  &  75.70  &  86.58  &  92.95  &  96.61    \\
&  \% edges sampled & 2.20  &  4.41  &  8.47  &  15.97  &  28.28  &  44.70  &  63.90  &  81.33  &  93.07    \\
&   avg edge count & 1.01  &  1.03  &  1.08  &  1.16  &  1.33  &  1.66  &  2.33  &  3.67  &  6.41    \\
&  avg node count & 1.06  &  1.14  &  1.30  &  1.61  &  2.26  &  3.55  &  6.23  &  11.60  &  22.32    \\

\hline \hline
\multirow{4}{*}{\begin{minipage}{0.1in}FB: UCSD\end{minipage}}     

&   \% nodes sampled & 5.73  &  11.12  &  20.30  &  33.69  &  51.16  &  67.93  &  81.37  &  90.38  &  95.93    \\
&  \% edges sampled & 1.92  &  3.93  &  7.77  &  14.21  &  25.91  &  42.10  &  61.34  &  79.05  &  91.90    \\
&   avg edge count & 1.02  &  1.03  &  1.07  &  1.13  &  1.28  &  1.59  &  2.19  &  3.38  &  5.83    \\
&  avg node count & 1.05  &  1.11  &  1.23  &  1.46  &  1.97  &  2.96  &  4.96  &  8.92  &  16.81    \\

\hline 
\end{tabular}}
\vspace{-5pt}
  \caption{
 Uniform sampling without replacement. Total \textbf{\% Nodes Sampled} in the graph when including all egos and neighbors. Total \textbf{\% Edges Sampled} in the graph when including all edges between egos and neighbors. \textbf{Average Edge Count}: ratio between all edges over unique edges. \textbf{Average Node Count}: ratio between all nodes over unique nodes.  }
  \label{tab:results}
\vspace{-25pt}
\end{table}

\section{Facebook}
\label{sec:facebook}

\subsection{Dataset Description}
In previous work \cite{gjoka10_walkingfb}, we collected a representative sample of $\approx 1$ million unique Facebook (FB) users by crawling the social graph using a Metropolis Hasting Random Walk (MHRW) method. Subsequently we collected the egonets for $36,628$ unique nodes that were randomly selected from the MHRW sample. This sub-sampling eliminates the correlation of consecutive nodes in the same crawl.  The representativeness of this data has been validated against true random samples from the Facebook taken during the same period \cite{gjoka10_walkingfb,gjoka2011practical}.  This sample closely approximates a uniform, with replacement sample of egonets from the publicly visible FB graph. 
In this sample all neighborhoods are uniquely labeled which allows for estimation using either the CDS or CC estimators. The value of the ``Average Edge Count'' is $1.03$ which according to our heuristic  in \Sec{sec:simulation} is an indication that the information provided by labels would not be very useful. Hence, we employ CDS estimators for our analysis.
We use the population size $N=240M$ which was estimated for this dataset by \cite{Katzir2011} and agrees with the  statistics reported by Facebook during the collection of the dataset (April 2009).

We complement this egonet sample with gender attributes for each user. We were able to fetch the publicly declared user-declared gender for 90\% of sampled users by crawling the url at \textit{\url{http://graph.facebook.com/userid}}. Additionally, we classified another 9.5\% by a majority rule that uses the first name of each user and a database of the number of times that first names were assigned to males and females. We first used the list of first names from the US Social Security records. If there was no match we then used the list of first names from the population of Facebook users with declared gender.
Last, we used \cite{genderPredictor} 
to predict the gender for the remaining 0.5\% users with a Naive Bayes classifier, based on the letter composition of first names.

\subsection{Results}

\begin{figure}
\centering
\subfigure{
\includegraphics[width=0.45\textwidth]{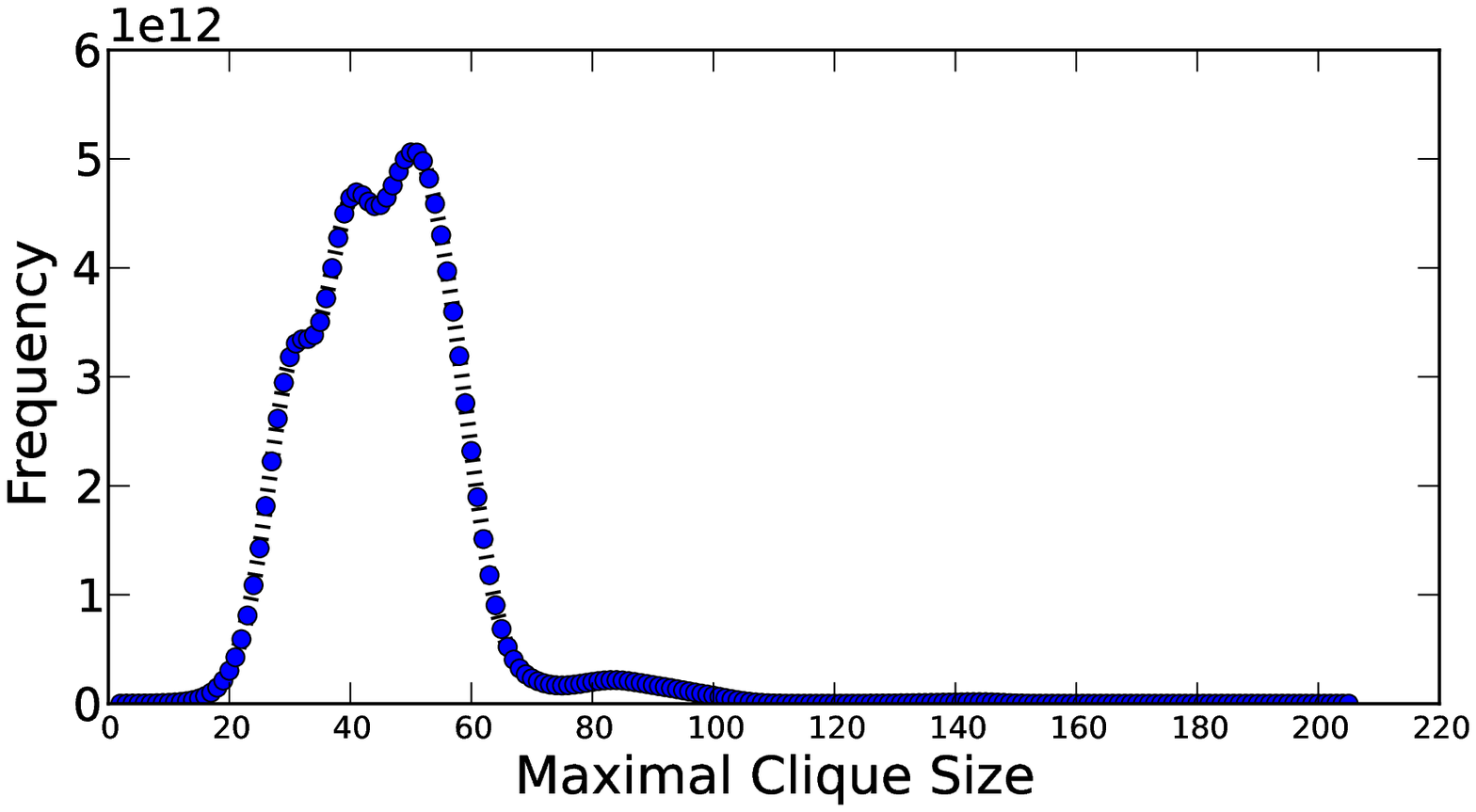}
}
\subfigure{
\includegraphics[width=0.45\textwidth]{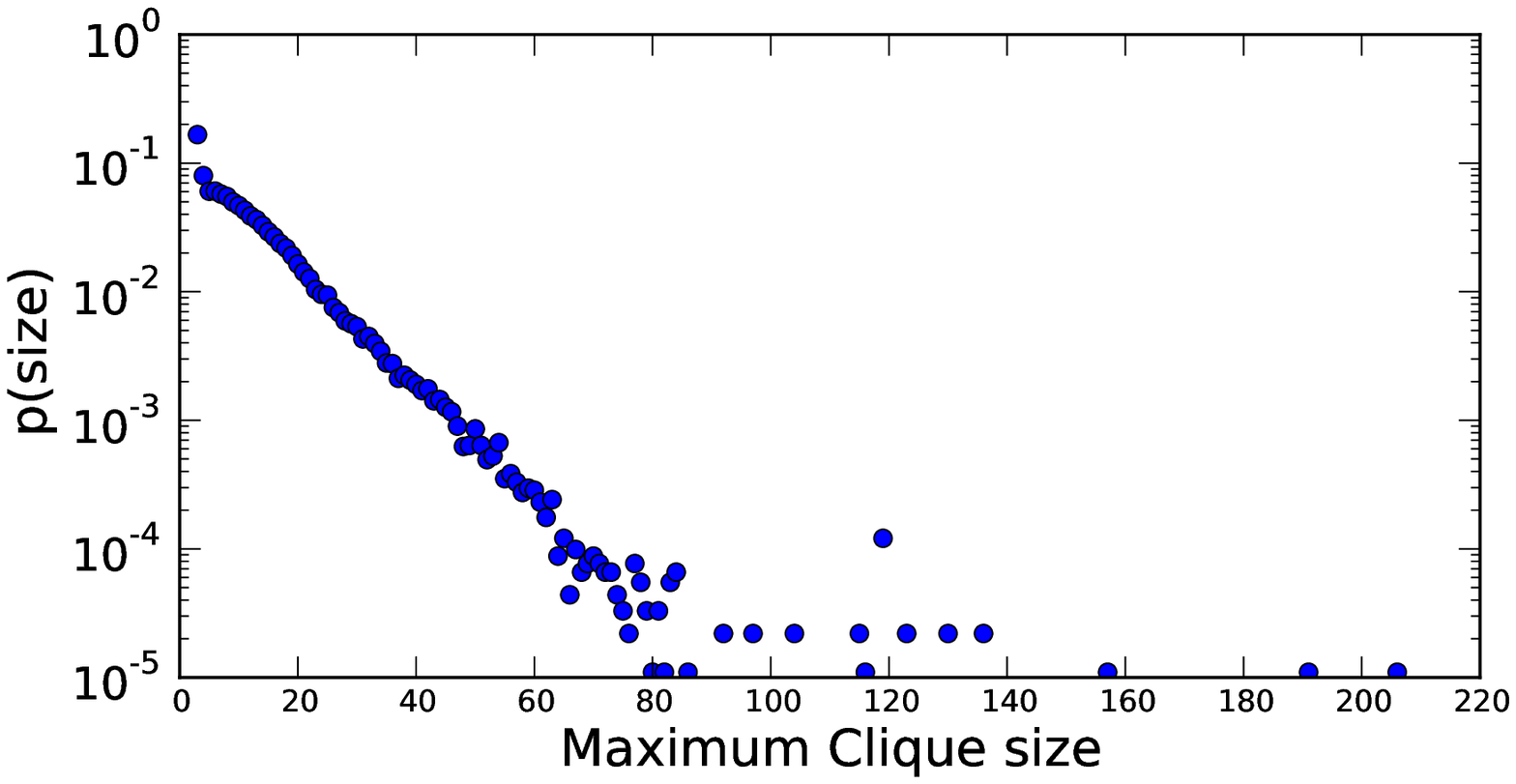}
}
\vspace{-5pt}
\caption{Estimated clique size distribution (Facebook social graph '09); top panel shows CDS estimates of maximal clique frequency, bottom panel shows the distribution of maximum clique sizes by ego.}
\label{fig:fbpublic_Ci}
\vspace{-5pt}
\end{figure}

\begin{figure*}
\centering
\includegraphics[width=1\textwidth]{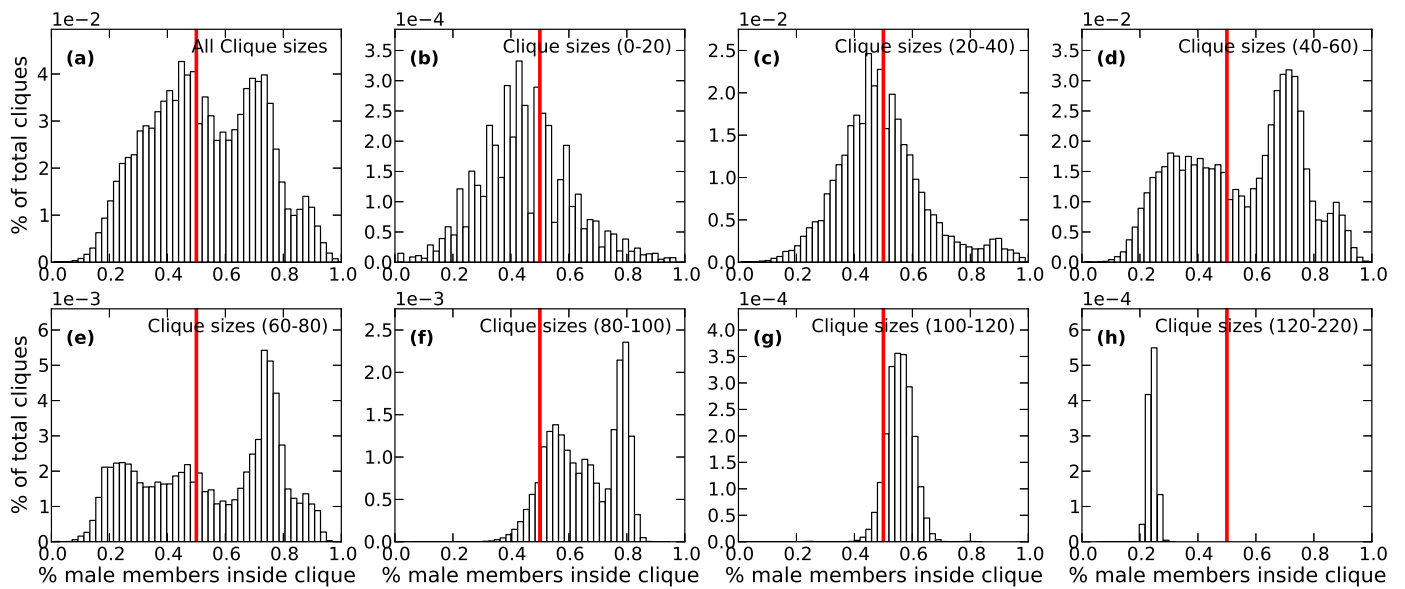}
\caption{CDS estimates for gender composition of maximal cliques, by order. (Facebook social graph '09)}
\vspace{-5pt}
\label{fig:fbpublic_Cu_every20}
\vspace{-5pt}
\end{figure*}

The top panel of \Fig{fig:fbpublic_Ci} shows the estimated distribution of maximal clique sizes over the entire FB social graph.  The FB graph is known to be highly clustered, and it indeed contains many large cliques: the modal clique size is 50, with the largest observed clique containing over 205 individuals.  Interestingly, the form of the clique distribution is neither monotone nor unimodal; significant peaks occur at 32, 41, and 50, with a minor mode near 84.  This suggests substantial heterogeneity in the mechanisms of clique formation, a point underscored by our findings regarding gender (see below).  

Rather more order is shown in the distribution of maximum clique sizes by ego (i.e., the largest clique to which a given individual belongs).  (\Fig{fig:fbpublic_Ci}, bottom panel.)  This shows a monotone distribution with a stable exponential decay over the range that is well-supported by our data.  Membership in moderate to large cliques is thus quite rare, despite their prevalence in the FB graph.

Beyond size distributions, our estimators allow us to examine how the composition of cliques varies across the FB graph.  \Fig{fig:fbpublic_Cu_every20} shows the estimated gender composition of FB cliques for all cliques (panel a) and cliques of varying order (panels b-h).  The $X$ axis in each panel indicates the fraction of clique members who are male, from 0 (entirely female) to 1 (entirely male); a vertical reference line indicates gender parity.  Our results provide clear evidence for strong heterogeneity in the makeup of FB cliques.  We see several distinct modes with characteristic gender frequencies, that occur over specific size ranges.  These include: a ``small equal clique'' mode of near-parity cliques of size 0--40; a 70--80\% male mode for cliques of size 40-100; a 60-80\% female mode for cliques of order 40--80; a second near-parity mode over the small range of sizes 100--120; and a strongly female dominated mode of very large cliques (sizes $>120$).  Although our data does not allow 
us to establish the mechanisms underlying these modes, we speculate that each is the result of a particular collection of social settings (e.g., fraternities or sororities, family groups, schools, or work organizations) that acts as a focus \cite{feld:ajs:1981} for tie formation.  Systematic variation in the gender composition of these settings then leads to corresponding variation in clique composition.  Interestingly, our findings do not corroborate the predictions of \cite{mayhew.et.al:sf:1995} regarding the relationship of clique size to gender homogeneity based on their analysis of face-to-face groups: while they posit a strongly negative relationship between heterogeneity and group size, we find that the FB graph supports a large fraction of near-parity cliques at even quite large orders.  While it is also true that extremely gender-homogeneous cliques become relatively more prevalent at large orders (versus small ones), the phenomenon appears to be uneven and size-specific.  Neither do we observe the 
power-law decay in group sizes reported by \cite{mayhew.et.al:sf:1995} for naturally occurring groups.  Since these prior results were based on observations of cliques in public, face-to-face settings, this lack of replication does not necessarily call into question the validity of the authors' conclusions in their original context; however, it does underscore that the formation of friendship cliques in OSNs may operate very differently from the sorts of groups examined in previous studies.

\section{Conclusion}
\label{sec:conclusion}
In this paper, we introduced novel unbiased estimators of clique composition and size distributions based on egocentric network samples. We presented two techniques, one of which exploits labeling of nodes (CC) and one which does not require this information (CDS). Both techniques are easily parallelizable, and suitable for use with large graphs.  (A Python implementation of our estimators is available at \cite{clique_estimators}.) We evaluated estimator performance via simulated sampling from real-world graphs, showing that both proposed techniques work well and that CC generally outperforms CDS as the sample ``saturates'' the graph; because CC imposes higher space complexity, we provided a simple heuristic that suggests when gains from using CC are likely to be substantial.  Last, we demonstrated an application of our estimators to clique composition in OSNs. We applied our methodology to egocentric samples collected in Facebook, which we complemented with gender information, allowing us to estimate the joint size and composition distribution of FB cliques with respect to gender. 
Our results underscore important differences between online and (previously reported) offline group structure, and provide evidence for strong gender heterogeneity in the makeup of FB cliques.

{
\bibliographystyle{abbrv}
\bibliography{related.bib}
}

\end{document}